\documentclass[%
 aip,
 jmp,%
 amsmath,amssymb,
 reprint
]{revtex4-1}

\usepackage{graphicx}
\usepackage{dcolumn}
\usepackage{bm}
\usepackage{natbib}
\usepackage{longtable}
\usepackage{booktabs}
\usepackage[table]{xcolor}
\usepackage{placeins}
\usepackage{textcomp}

\let\centering\relax
\begin{document}


\title[Anharmonic cascade IR spectra of PAHs]{Fully anharmonic infrared cascade spectra of polycyclic aromatic hydrocarbons}

\author{Cameron J.~Mackie}
 \email{mackie@lbl.gov}
 \affiliation{Chemical Sciences Division, Lawrence Berkeley National Laboratory, Berkeley, California 94720, USA}
 \affiliation{Leiden Observatory, Leiden University, PO Box 9513, 2300 RA Leiden, The Netherlands}
\author{Tao Chen}
 \affiliation{Leiden Observatory, Leiden University, PO Box 9513, 2300 RA Leiden, The Netherlands}
 \affiliation{Department of Theoretical Chemistry and Biology, School of Biotechnology, Royal Institute of Technology, 10691, Stockholm, Sweden}
\author{Alessandra Candian}
 \affiliation{Leiden Observatory, Leiden University, PO Box 9513, 2300 RA Leiden, The Netherlands}
\author{Timothy J.~Lee}
 \affiliation{NASA Ames Research Center, Moffett Field, California 94035-1000, United States}
\author{Alexander G.~G.~M.~Tielens}
 \affiliation{Leiden Observatory, Leiden University, PO Box 9513, 2300 RA Leiden, The Netherlands}

\date{\today}

\begin{abstract}
The infrared (IR) emission of polycyclic aromatic hydrocarbons (PAHs) permeates our universe; astronomers have detected the IR signatures of PAHs around many interstellar objects. The IR emission of interstellar PAHs differs from their emission as seen under conditions on Earth, as they emit through a collisionless cascade down through their excited vibrational states from high internal energies. The difficulty in reproducing interstellar conditions in the laboratory results in a reliance on theoretical techniques. However, the size and complexity of PAHs requires careful consideration when producing the theoretical spectra. In this work we outline the theoretical methods necessary to lead to a fully theoretical IR cascade spectra of PAHs including: an anharmonic second order vibrational perturbation theory (VPT2) treatment; the inclusion of Fermi resonances through polyads; and the calculation of anharmonic temperature band shifts and broadenings (including resonances) through a Wang--Landau approach. We also suggest a simplified scheme to calculate vibrational emission spectra that retains the essential characteristics of the full IR cascade treatment and can directly transform low temperature absorption spectra in IR cascade spectra. Additionally we show that past astronomical models were in error in assuming a 15 cm$^{-1}$ correction was needed to account for anharmonic emission effects.
\end{abstract}

\maketitle

\section{\label{sec:intro}Introduction}
Polycyclic aromatic hydrocarbons (PAHs) are a family of molecules characterized as containing multiple fused benzenoid rings with the free edges capped with hydrogen. Their size, complexity, stability, and universal abundance make their study appealing to vastly different areas of research: from biology for their carcinogenic properties\cite{pah_cancer}, to material science for their desirable structural properties\cite{Lai200123, doi:10.1021/jp2095032, doi:10.1021/cm301993z}, to engineering for their role in the undesirable combustion byproducts of engines and rockets\cite{doi:10.1021/es001028d}. Perhaps the most surprising field to study PAHs is astronomy. The infrared (IR) signatures of aromatic carbon--containing molecules have been observed in all interstellar objects with sufficient ultra--violet (UV) radiation to excite them; from reflection nebulae, to photo--dissociation regions, to emissions from distant galaxies as a whole\cite{2008ARA&A..46..289T}. These so--called ``aromatic infrared bands'' (AIBs) are generally attributed to PAHs and PAH derivatives\cite{1984A&A...137L...5L, 1985ApJ...290L..25A}. These interstellar PAHs absorb UV photons from a source (i.e., a parent star) becoming electronically excited, then quickly (hundreds of femtoseconds) return to their electronic ground state through emission-less internal conversion into excited vibrational states. They then relax radiatively through gradual ($\sim$ seconds) emission of IR photons from their excited vibrational states producing what is known as an IR cascade spectrum\cite{1987ApJ...315L..61B,1989ApJS...71..733A}. This process proceeds under non--thermal equilibrium conditions, that is to say each type of degree of freedom (vibration, electronic, translation, and rotation) is decoupled, giving rise to independent temperatures. A PAH will typically absorb a UV photon once a day and emit the absorbed energy through a vibrational cascade in about 1 second. For comparison, the collisional time scale is also of the order of a day. Hence, the emission process has to be evaluated as a microcanonical ensemble. The electronic, translational, and rotational temperatures of interstellar PAHs can be considered negligible due to the fast inter--conversion of electronic energy, low collision rates, and inefficient means of pumping rotational levels respectively. This leaves only the vibrational temperature as relevant. However, as radiative cooling is a stochastic process, each excited molecule will follow an individual trajectory through phase space. The cascade emission spectrum is then produced as an average over these different pathways.

Producing experimental IR cascade spectra under astrophysically relevant conditions (i.e., an extreme vacuum, with low collision rates, and quenched rotations) is extremely difficult with current technologies. Therefore, extrapolations from the available data must be made. Experiments have been limited largely to either high--temperature gas--phase absorption spectra\cite{:/content/aip/journal/jcp/106/22/10.1063/1.474033, Cane19971839, nist}, or low--temperature matrix--isolation spectra\cite{1994ASPC...58..283H, 1997JPCA..101.3472H, tet_matrix, 1998JPCA..102..344H, doi:10.1021/jp021938c, Mattioda2005156}. The usefulness of these experiments in modeling IR cascade spectra are complicated by temperature and pressure effects in the former, and unpredictable matrix interaction effects in the later. Recently, high--resolution, low--temperature, gas--phase \emph{absorption} spectra have been produced for a handful of small PAH species in the CH--stretching IR region and this has provided insight into the complexity of PAH IR spectra\cite{elena2015, elena2016, elena2017}.

Due to the large variety and size of PAHs, most theoretical calculations to date have been performed using low levels of theory, typically scaled density functional theory (DFT) calculations at the double harmonic level, using the B3LYP functional\cite{becke1993becke} and 4-31G\cite{hehre1972self} basis set. Databases of the theoretical calculations of PAHs have been compiled\cite{Malloci2007353, 2014ApJS..211....8B} and are used by astronomers in modeling PAH IR cascade emission. These databases have been vital to understanding the variation in the PAH emission observed both between different interstellar environments, and spatially within a given interstellar object. These variations can be used to track properties such as size, structure, temperature, and charge balance\cite{ricca2012infrared,hony2001ch,verstraete1990ionization}. 

Although successful in many ways, detailed comparison between theory and experiments have revealed that these double harmonic calculations cannot model accurately PAH spectra at high resolution, especially in the CH--stretching region centered around 3.3 $\mu$m\cite{elena2015, elena2016, elena2017}. While this has not hindered analysis of observations of PAHs using current telescopes, it will become a glaring issue with the upcoming launch of the James Webb Space Telescope, which will give an unprecedented combination of spectral and spatial resolution in the IR. In order to overcome these deficiencies an anharmonic approach is necessary. The anharmonic approach is not only superior in predicting the vibrational fundamental band positions and intensities themselves, but also allows for the calculation of combination bands, overtones, and hot--bands, key features for vibrationally excited PAHs. Additionally, anharmonic calculations can account for mixing between modes in \emph{Fermi resonances}, which are vitally important for reproducing the 3.3 $\mu$m region accurately.

Most important for the discussion here, these anharmonic calculations also allow for an ab--initio approach for generating internal energy/temperature--dependent spectra. As these anharmonic interactions create band shifts and broadening of the emission bands, accurate and efficient methods are required\cite{doi:10.1021/jp071610p, joblin1995infrared, cook1998simulated, pech2002profiles, joblin2011modeling}. Two main methods have been put forward: a direct approach -- whereby the thermal population of each vibrational state is calculated, then transition energies and intensities are calculated for every possible transition between populated states\cite{B814037E}; and a statistical approach -- whereby a biased Monte Carlo walk, known as the Wang--Landau method\cite{wang2001efficient, landau2004new}, is used to construct the density of states (DOS) and the energy--dependent spectra. These methods then produce temperature--dependent spectra through a Laplace transformation\cite{basire2009temperature}. The latter method will be shown in this work to be advantageous in that incorporating the important Fermi resonances into the spectra is straightforward and can be made computationally efficient. 

Further, the collision--free cascade emission process of PAHs needs to be considered when modelling the IR spectrum of interstellar PAHs. As stated previously, the PAHs absorb UV photons, the resulting electronic excitation is then quickly interconverted into highly excited vibrational states of the ground electronic state, then the energy is emitted through IR photons as they cool. 

It is important to note that the PAH molecules present in interstellar environments are large and contain many, many vibrational degrees of freedom; so many that even with 10--12 eV of excess vibrational energy, no single vibrational degree of freedom will be highly excited. Moreover, PAH molecules are very rigid structures.  These two features mean that the 10--12 eV excited PAHs are ideally suited to be treated by second--order vibrational perturbation theory (VPT2), with proper treatment of resonance polyads. 

Attempts have been made to replicate astrophysical conditions in the laboratory\cite{cherchneff1989infrared, brenner1992infrared, kim2001single} with some success. Theoretical modelling of the IR cascade emission has also been performed, typically using DFT calculations at the double harmonic level\cite{cook1998simulated, pech2002profiles, Candian11042015}, and to a lesser extent at the anharmonic level\cite{joblin2011modeling,PARNEIX2012112}. However, a sufficient treatment of Fermi resonances in a full anharmonic cascade model has never been taken into account previously, leading to spectra that could not be fully trusted, especially in the troublesome 3.3 $\mu$m region. Additionally, due to the limited datasets and number of species analyzed, these previous studies have given rise to methods whose validity has to be assessed. 

With the proper incorporation of anharmonicities, resonances, and internal energy/temperature dependence (including the incorporation of polyads), the ingredients are present for a full theoretical anharmonic IR cascade spectra of PAHs. In this work, these calculations are performed for a small subset of astrophysically relevant species, including regular, hydrogenated, and methylated PAHs, as well as cationic species (see supporting material for all results).

\section{\label{sec:theory}Theory}
A brief summary of each theoretical method used in this work is given below. For full descriptions refer to cited works.

\subsection{\label{sub:anharm}Anharmonicity}
In the double harmonic approximation the vibrational potential of a molecule in its stationary point geometry is given by

\begin{equation}
\label{harmonic}
V = \frac{1}{2}\sum_{i,j}^{3N} F_{ij} X_i X_j
\end{equation}

where $V$ is the total vibrational potential, $F_{ij}=\frac{\partial V}{\partial X_i \partial X_j}$ are the quadratic force constants between atoms i and j, and X$_i$ and X$_j$ are the Cartesian coordinates of atom $i$ and $j$ relative to the stationary point geometry.

The quadratic force constants are computed through ab--initio analytical second derivative techniques. Once computed, these quadratic force constants are used to construct a Hessian matrix, which upon diagonalization yields the individual vibrational band energies (harmonic frequencies) as the eigenvalues, and the individual vibrational mode displacement descriptions or ``normal modes'' as the eigenvectors. The harmonic intensities of the vibrational modes are calculated using the double harmonic approximation, meaning that the vibrational wavefunctions are those from the harmonic oscillator approximation and the dipole operator is truncated at the first derivative. This approximation however, contains a number of short-comings: frequencies are too high in energy; intensities of overtones, combination bands, and hot bands are exactly zero; and interactions between modes, such as couplings and resonances, cannot be taken into account. As a result, empirical corrections (i.e., scaling factors\cite{langhoff1996theoretical}) and empirical absorption and emission models (i.e., temperature effects\cite{joblin1995infrared,cook1998simulated}) are applied in order to attempt to produce spectra that agree better with experiment. While largely successful in reproducing the overall IR features of PAHs, the finer details are absent. In order to account for these short comings, a more thorough theoretical approach is necessary. 

Specifically, the atomic potential needs to be represented in a manner that is closer to the physical reality; moving away from the harmonic potential towards an anharmonic potential. Ideally, the exact potential could be written as an infinite order Taylor series. (Equation \ref{harmonic} above would represent the second order terms of such an expansion.) Subsequent higher order terms making use of cubic, quartic, pentic, hextic, etc. force constants can be included. In order to sufficiently account for combination bands, overtones, hot--bands, couplings, resonances, and temperature effects it is often enough to include the first three non--zero terms of the expansion, leading to the so--called anharmonic quartic force field (QFF).

\begin{equation}
\label{taylor}
\begin{aligned}
V &= \frac{1}{2}\sum_{i,j}^{3N} \left(\frac{\partial^2 V}{\partial X_i \partial X_j}\right) X_i X_j \\
 &+ \frac{1}{6}\sum_{i,j,k}^{3N} \left(\frac{\partial^3 V}{\partial X_i \partial X_j \partial X_k}\right) X_i X_j X_k \\
 &+ \frac{1}{24}\sum_{i,j,k,l}^{3N} \left(\frac{\partial^4 V}{\partial X_i \partial X_j \partial X_k \partial X_l}\right) X_i X_j X_k X_l\\
\end{aligned}
\end{equation}

A simple Hessian can no longer be constructed to solve this potential since it is now non--linear. Instead, the harmonic solution is produced as usual and the cubic and quartic force constants are used to perturb the solution using a vibrational second--order perturbation theory treatment (VPT2) (see references \citenum{csaszar2012anharmonic}, \citenum{aliev1985molecular}, and \citenum{papouvsek1982molecular}). The total energy of a specific vibrational state is given by (asymmetric top case)

\begin{equation}
\label{perturb}
\begin{aligned}
E(n) &= \sum_k \omega_k \left(n_k + \frac{1}{2}\right) + \sum_{k\le l} \chi_{kl}\left(n_k + \frac{1}{2}\right) \\
&\times\left(n_l + \frac{1}{2}\right)
\end{aligned}
\end{equation}

where $\omega_k$ is the harmonic frequency of the $k^{th}$ fundamental vibrational mode, $n_k$ are the number of quanta in the $k^{th}$ fundamental vibrational mode, and $\chi_{kl}$ are the corresponding anharmonic constants. The anharmonic constants are calculated using the cubic and quartic force constants (see Refs. \citenum{aliev1985molecular,papouvsek1982molecular,Burcl20031881}). The energies of the combination bands, hot--bands and overtones are determined by increasing the number of quanta in the relevant vibrational modes and taking the difference between the two states. Equation \ref{perturb} also shows that the vibrational band position of a ``fundamental'' will be slightly different if ``spectator'' modes have n $>$ 0, or in other words, if the molecule is not in the ground vibrational state. Equation \ref{perturb} is a valid approximation for the vibrational state energies as long as the number of quanta in any given mode is low. Higher excitations would require additional force constants to be included in higher order perturbation theory approaches. 

Intensities can also be calculated at the anharmonic level through the use of VPT2 including fundamentals, overtones, and double combination bands. The equations are quite extensive, see Ref. \citenum{gaussian_anharm_I}. 

\subsection{\label{sub:polyads}Resonances and polyads}
In an anharmonic treatment the vibrational modes are not independent from one another. An important manner in which they can interact is through \emph{resonances}. A resonance occurs when the energy of the normal modes or the sum of modes are close to one another in energy. Mathematically speaking, a resonance occurs in the context of a VPT2 treatment when the denominator of a term in the anharmonic constant calculation approaches zero, causing the offending term to increase dramatically in value. These singularities or resonances are removed from the VPT2 treatment and handled separately in a method akin to second order degenerate perturbation theory (see Refs. \citenum{aliev1985molecular,papouvsek1982molecular,doi:10.1080/00268978900100751,martin1997accurate}, and references therein). In a typical VPT2 treatment four main types of resonance occur: Darling--Dennison, when two overtones or combination bands are approximately equal in energy; type one Fermi, when a fundamental vibration is approximately equal to the energy of an overtone; type two Fermi, when one fundamental vibration is approximately equal to the sum of two other vibrations; and 1:1 resonances, when two fundamental vibrations are approximately equal\cite{martin1997accurate}. The effect of a resonance on a spectrum is to push the bands of the vibrational states involved apart in frequency (compared to their non--resonant frequencies) while the sum of their intensities is redistributed (unequally) between them. 

Complicating matters further, any given vibrational state can be simultaneously involved in many separate resonances. It is not sufficient to handle each resonance individually, instead these groups of mutually resonating states need to be computed together. A resonance polyad matrix is constructed, with the diagonal terms being the perturbed states as given in equation \ref{perturb} (with the offending resonant terms removed from the anharmonic constants) and the off--diagonal terms being coupling constants (see Ref. \citenum{aliev1985molecular,papouvsek1982molecular,martin1997accurate, martin1995anharmonic, doi:10.1080/00268978900100751} for more details). Diagonalization of this matrix results in new vibrational band positions as the eigenvalues, and the strengths of the resonant mixings as the square of the individual components of the eigenvectors. These resonant mixing terms can be used to determine the redistribution of intensities between the resonating modes. This treatment has been previously applied to a series of PAHs\cite{mackie2015b, mackie2016anharmonic, mackie2017} and was shown to be crucial in reproducing the CH--stretching region due to a large number of type two Fermi resonances.

\subsection{\label{sub:temp}Temperature--dependent spectra}
Theoretical approaches in addressing the temperature effects of the IR spectra of PAHs has been tackled by a number of papers, see Refs. \citenum{basire2008quantum,basire2009temperature,B814037E,calvo2011temperature,joblin2011modeling}. Producing temperature--dependent spectra from a QFF using a Wang--Landau style walk has been shown previously to work successfully for PAHs\cite{basire2009temperature,calvo2011temperature}, therefore similar methods have been adopted in this work, following the process as outlined in Ref. \citenum{basire2008quantum, basire2009temperature}. 

The initial step towards producing a temperature--dependent spectrum is to calculate the vibrational density of states (DOS). As the number of atoms in a molecule increases, the number of possible states falling in a given energy range grow extremely fast, especially at high internal energies. For this reason exact counting of the DOS becomes impossible. Therefore, Monte Carlo type samplings are typically used, in particular the Wang--Landau method\cite{wang2001efficient,landau2004new}. 

The aim of a Wang--Landau walk is to visit all predefined energy bins ($\delta E$) equally in an efficient manner. This is accomplished through biasing the walk to prefer states that have been visited fewer times than the energy bin of the current state. Initially, the DOS ($\Omega$) is set to 1 for the whole energy range of interest ([E$_{min}$,E$_{max}$]). The set of vibrational quantum numbers \{n\} is initially set to a random state inside the given energy range. At each step there is a probability that any of the vibrational quantum numbers can increase by one quanta, decrease by one quanta, or remain the same. The value of this probability is chosen to be large enough so that the walk covers the whole energy range in a reasonable amount of steps, but also small enough as to not jump outside of the energy range constantly. Typical values for this probability range from 4--23\%. After the set of new quantum numbers, \{n\}$_{new}$, are generated the probability (P) that this state is accepted into the accumulation of $\Omega$ is given by

\begin{equation}
\label{wang}
P(\{n\}_{current} \rightarrow \{n\}_{new}) = min[1,\frac{\Omega(E_{current})}{\Omega(E_{new})}]
\end{equation}

If the new state is accepted then $\Omega$(E$_{new}$) is updated by adding a quantity $f$, where $f$ is a modification factor set to some initial quantity greater than zero (typically the mathematical constant $e$). Since $\Omega$ can become quite large, the value ln($\Omega$) is generally stored. A histogram (H) of the visits to each energy bin is accumulated in order to test for equal sampling of the walk. After sufficient flatness in the sampling is achieved, the value of $f$ is updated by taking its square root, H is reset, and the walk is continued again until flatness is once again achieved. This process of updating $f$ and re--running the walk ensures quick and accurate convergence towards the true DOS. Typically, $f$ is updated and the walk re--run on the order of 20 times, depending on the desired accuracy of the DOS.

Next, an energy--dependent spectrum is produced. A separate Wang--Landau walk over the same internal energy range is performed. With the acceptance criteria of a state (P) now being based on the completed $\Omega$(E) values. At each step of this second walk an accumulated spectrum, $\mathcal{S}$, at internal vibrational energy E is updated according to

\begin{equation}
\label{energy}
\mathcal{S}(\nu,E) = \sum_k(n_k+1)\sigma^{(k)}_{0\rightarrow 1}\delta[h\nu - \Delta E^{(k)}]
\end{equation} 
with $\sigma^{(k)}_{0\rightarrow 1}$ being the absorption cross section of mode k or combination band at the vibrational ground state. For combination bands n$_k$ is taken to be the smaller of the two vibrational quantum numbers of the two bands involved.

The energy difference at the anharmonic level for a given vibrational transition of mode k ($n_{k}$ $\rightarrow$ $n_{k}+1$) is then given by

\begin{equation}
\label{deltaE}
 \Delta E^{(k)}(\{n\}) = \omega_k + 2\chi_{kk}(n_k + 1) + \frac{1}{2}\sum_{i \neq k} \chi_{ik} +  \sum_{i \neq k} \chi_{ik}n_i
\end{equation}

Conversely, if the accumulated \emph{emission} spectrum ($n_{k}$ $\rightarrow$ $n_{k}-1$) is desired, then equation \ref{energy} is replaced with 
\begin{equation}
\label{energy_em}
\mathcal{S^\prime}(\nu,E) = \sum_k(n_k)\sigma^{(k)}_{1\rightarrow 0}\delta[h\nu - \Delta E^{\prime(k)}]
\end{equation} 

and equation \ref{deltaE} with

\begin{equation}
\label{deltaE_em}
 \Delta E^{\prime(k)}(\{n\}) = \omega_k + 2\chi_{kk}(n_k) + \frac{1}{2}\sum_{i \neq k} \chi_{ik} +  \sum_{i \neq k} \chi_{ik}n_i
\end{equation}

Once the walk is complete, the states are then normalized by the number of visits to the corresponding energy in the histogram $\mathcal{N}$, giving the microcanonical absorption (emission) intensity 

\begin{equation}
I(\nu,E) = \frac{\mathcal{S}(\nu,E)}{\mathcal{N}(\nu,E)}
\end{equation}

While astronomers are interested in (microcanonical) spectra evaluated as a function of internal energy, laboratory spectra are often measured at constant temperature. The (canonical) intensity spectrum follows from a Laplace transformation

\begin{equation}
I(\nu,T) = \frac{1}{Z(T)}\int_{E_{min}}^{E_{max}} I(\nu,E)\Omega(E)e^{(-E/k_BT)}dE
\end{equation} 

With Z being the partition function
\begin{equation}
Z(T) = \int \Omega(E)e^{(-E/k_BT)}dE
\end{equation}

The maximum and minimum reliable T values are determined by the energy range of $\Omega$. $\Omega(E)e^{(-E/k_BT)}$ at a given T should die off rapidly at E$_{min}$ and E$_{max}$ to ensure the integral over E$_{min}$ to E$_{max}$ is close to the true integral over the infinite energy range. It should also be noted that as the number of quanta in a given vibrational mode increases the accuracy of the anharmonic constants decreases (i.e., the anharmonic spacing of the vibrational energy levels are decreasing erroneously at a constant rate). Fortunately for PAHs, even with high \emph{total} internal vibrational energy (e.g., 12 eV), there are on average less than one quanta per vibrational mode. 

\subsection{\label{sub:polytemp}Polyads and temperature dependence}
\begin{figure}
\resizebox{\hsize}{!}{\includegraphics{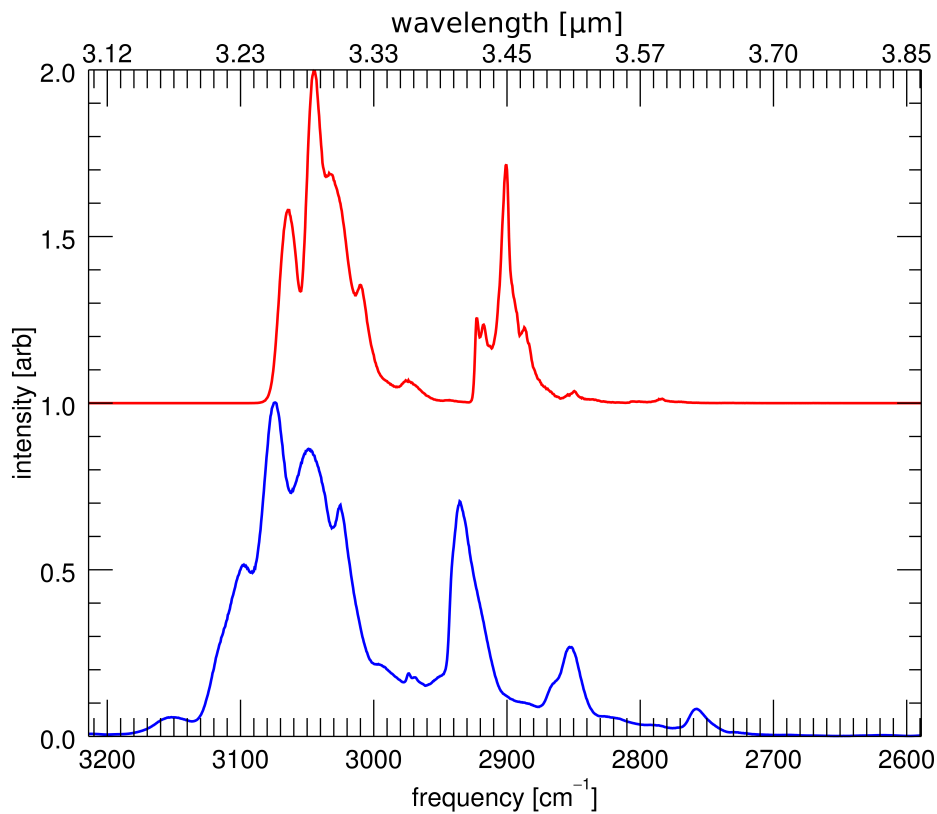}}
\caption{The effect of including polyads (blue, bottom panel) and excluding polyads (red, top panel) from the anharmonic temperature--dependent calculations of 9--methylanthracene using the Wang--Landau approach.}
\label{fig:temp_poly_nopoly}
\end{figure}

\begin{figure}
\resizebox{\hsize}{!}{\includegraphics{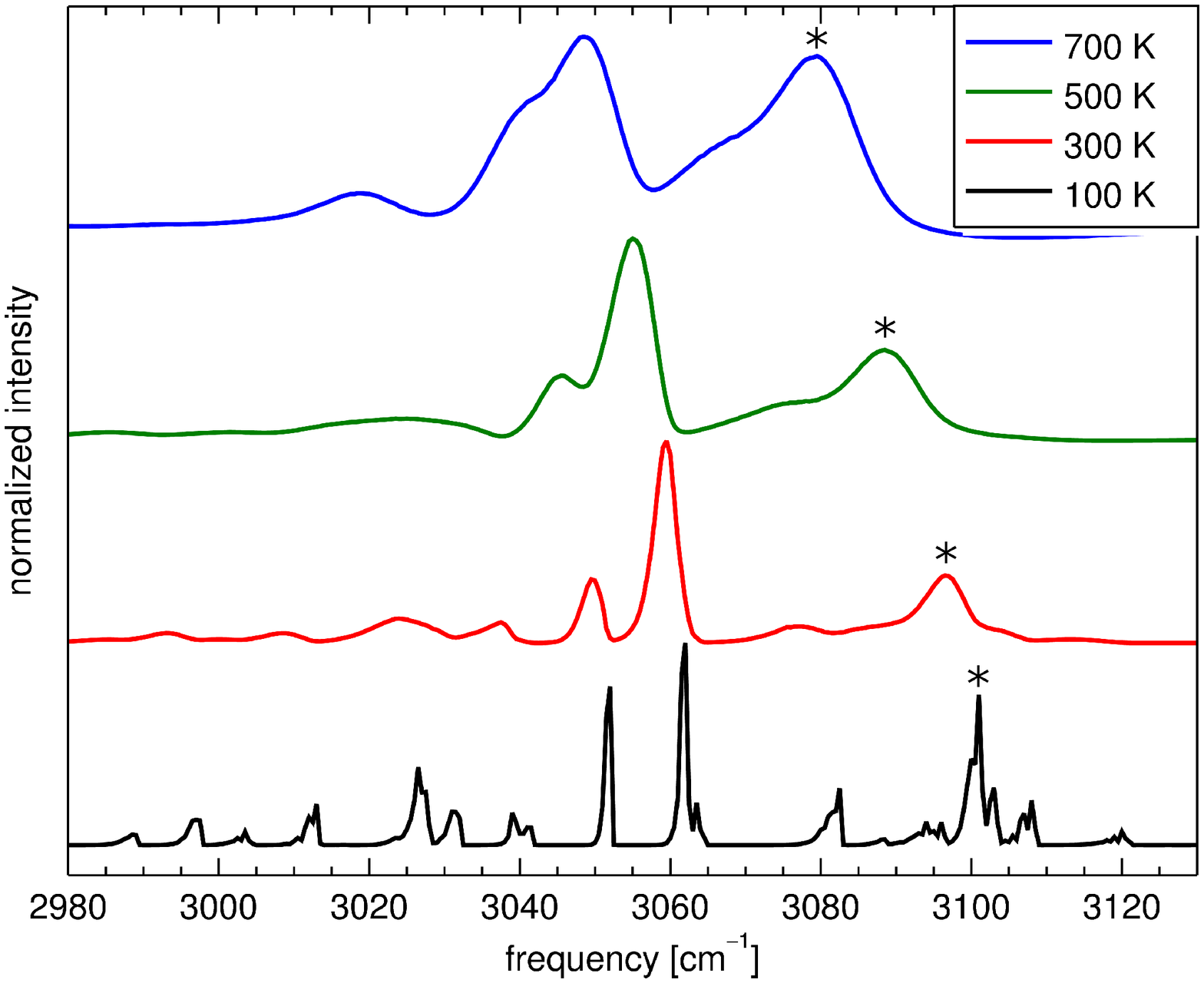}}
\caption{The theoretical infrared spectrum of tetracene showing an increase in relative intensity of a combination band (marked with a ``*'') as the internal temperature is increased.}
\label{fig:resoreso}
\end{figure}

Resonances play a dominant role in the IR spectra of PAHs, especially in the CH--stretching region (3200 -- 2950 cm$^{-1}$, 3.3$\mu$m). Unfortunately, their calculation is a significant bottle neck in the computation of even non--temperature--dependent spectra, and the introduction of temperature dependence can mean recalculating and diagonalizing these resonance polyad matrices millions of times, depending on the desired resolution and maximum temperature. Therefore, an efficient handling of resonances is crucial.

To limit the size of the polyads and, as a result, perform the diagonalizations in a computationally efficient manner we have separated the polyads by symmetry and energy separation. This decreases the computational load in a number of ways: 1) Since the vibrational states of differing symmetry do not couple in the polyad (i.e., the polyad is symmetry blocked), we can safely separate the full polyad into individual polyads only containing the vibrational states of the same symmetry. 2) We can completely avoid the calculation of polyads containing states with zero intensity based on symmetry. 3) We do not need to determine a--priori the relevant resonances based on thermal populations, (i.e., we do not have to use other methods to shrink our polyads). 4) The naturally occurring energy gaps between vibrational mode types allow for the polyads to be further separated, since for example, a vibrational mode at 800 cm$^{-1}$ will not have a significant coupling with one at 3000 cm$^{-1}$ (discussed in more detail below).

With these considerations in mind, the incorporation of polyads into a Wang--Landau treatment of the IR energy/temperature--dependent spectrum becomes straight forward. During the generation of a spectrum, for each iteration of the Wang-Landau method the polyad matrices are reconstructed using the ground state (symmetry blocked) resonance matrices with the diagonal terms being updated by the new transition energies given by equation \ref{deltaE}, or in the case of combination bands and overtones the sum of the transition energies of the modes involved. The off--diagonal coupling terms of the resonance matrix are left unchanged as they do not depend on the vibrational energy levels occupied, but only the change in the occupation of vibrational levels of the transitions ($\Delta$(n$_{k}$) is always 1). Once modified, the eigenvalues of the altered resonance matrix give the new transition energies of the vibrationally excited states (i.e., a new set of $\Delta$ E$^{(k)}$(\{n\}), and the squared coefficients of the eigenvalues give the intensity sharing percentages over the resonating modes (i.e., a new set of $\sigma^{(k)}_{0\rightarrow 1}$). All other steps during the Wang--Landau creation of the temperature--dependent spectrum remain the same\cite{basire2009temperature}.

Figure \ref{fig:temp_poly_nopoly} shows the effect of incorporating polyads into the Wang--Landau calculation of the temperature--dependent spectrum of 9-methylanthracene at 1000 Kelvin. In the top spectrum (red) polyads are ignored, while in the bottom spectrum (blue) polyads are included. Even with the line blending which occurs at high temperatures, discrepancies in band positions and widths, and the loss of features are all present when resonances are excluded.

It was found that the resonance between a particular combination band and a fundamental can wax or wane as thermal population of vibrational levels increase. Since the strength of the interactions between resonating states depends on the energy separation of the states, and since the anharmonic constants differ between fundamentals and combination bands, the combination bands can ``catch--up'' in position and begin to borrow more intensity from the fundamentals. This has the effect of a gradual increase in intensity of the combination band as the internal energy of the molecules increases. For example, this effect was found to be very strong for a particular combination band in tetracene. The combination band was observed to gain significant intensity at higher energies. Figure \ref{fig:resoreso} shows the effect this has on the temperature--dependent spectrum. It can be seen that the combination band marked with a ``*'' initially drops in intensity (due to thermal broadening), but then grows again at higher temperatures compared to the two strongest fundamentals.

\subsection{\label{sub:cascade}PAH IR cascade spectra}
Interstellar PAHs are electronically excited by UV photons originating from stellar sources, then convert quickly to their ground electronic state through emissionless internal conversion to highly excited vibrational states. The PAHs then slowly emit this vibrational energy as IR photons, but much faster than a subsequent UV excitation process. When a PAH emits these IR photons, its internal energy is decreased by the energy of said photon. This decrease in internal energy then affects the energy at which the next IR photons emit (see equation \ref{deltaE}). This results in an ever changing IR emission spectrum during the cooling process. In the interstellar medium this process is stochastic, in that PAHs have time to cool fully to their vibrational ground states in between UV photon absorption events. Additionally, due to the low densities, the process is collisionless. This results in non--equilibrium conditions and has a significant effect on the resulting spectrum compared to laboratory measurements.

Modelling this process is straight forward, and has been described in Ref. \citenum{cook1998simulated, pech2002profiles, joblin2011modeling}. However, here we have fully incorporated resonances into the process as described above. The probability of a PAH absorbing a stellar UV photon is proportional to the product of the blackbody spectrum of a star of a given temperature and the UV absorption cross section of the PAH. The total internal vibrational energy E$_{tot}$ of the PAH is then equal to the energy of the absorbed UV photon. The probability of emitting an IR photon (E$_{IR}$) at a given internal energy (E$_{tot}$) is proportional to its rate, which in turn is proportional to the energy--dependent emission spectrum at E$_{tot}$ times the square of the corresponding frequencies. To model this, an IR photon is randomly selected based on these probabilities and a histogram of photon counts is updated at the corresponding energy of E$_{IR}$. The total internal energy contained in the PAH is then updated as: E$_{tot_{new}}$ = E$_{tot}$ - E$_{IR}$. The corresponding energy is then used to look up the new energy--dependent emission spectrum at E$_{tot_{new}}$, giving a new set of emission probabilities at the lower energy. This cascade process is repeated until the PAH has fully cooled down to its vibrational ground state. At this point a new UV photon is absorbed and the full cascade is repeated again. This process continues until the desired accuracy is achieved.
 
\section{\label{sec:method}Implementation}
\subsection{\label{sub:qff}Quartic force fields}
The QFFs of the PAHs in this work were calculated using Gaussian 09\cite{GAUSSIAN..09} within the DFT framework, using the B3LYP functional\cite{becke1993becke} and the N07D basis set\cite{barone2008development}. A tight convergence criterion was used for the geometry optimization and a very fine grid (Int=200974) was used for numerical integrations. 

\subsection{\label{sub:VPT2}VPT2}
The VPT2 treatment was handled by a locally modified version of SPECTRO\cite{spectro}. SPECTRO has the flexibility to handle the polyads as described in section \ref{sub:polyads}. Additionally, the SPECTRO output of the resonant matrices facilitated the energy/temperature incorporation as described in section \ref{sub:polytemp}. For more information on how polyads are handled in SPECTRO, see Ref. \citenum{martin1997accurate}.  

\subsection{\label{sub:polyads}Polyads}
The cut--off for considering states to be in resonance is controlled by two parameters in SPECTRO, the minimum interaction strength between the two bands (W), and the difference in energy between the two bands ($\Delta$). The default value of $\Delta$ (200 cm$^{-1}$) has been shown to reproduce the CH--stretching region of PAHs accurately\cite{mackie2016anharmonic, mackie2017}. However, polyad sizes can grow quite large with this cut--off, especially when the molecule has a low symmetry. For example, the three IR active polyads of benz[a]anthracene (C$_{18}$H$_{12}$, C$_s$ symmetry) contain 117, 675, and 204 states respectively. This typically occurs because of \emph{resonance chaining}, where a series of modes resonate with their neighbors who in turn resonate with their neighbors etc., leading to a polyad where the highest energy mode and the lowest energy mode could end up with $\Delta$ $>$ 1500 cm$^{-1}$. It was found that the CH--stretching region (above 3000 cm$^{-1}$) is extremely sensitive to the lowering of $\Delta$ due to strong Fermi resonances and was thus left alone. The vibrational modes below 2000 cm$^{-1}$ were found to only weakly interact through Fermi resonances, and as a result lowering $\Delta$ to 25 cm$^{-1}$ showed little effect on the resulting spectra. Fortunately, a naturally large break occurs between the CH--stretching vibrational modes and all other vibrational modes so no resonance chaining occurs between these two spectral ranges. In order to break these chains and further shrink the polyad sizes, two separate $\Delta$ values were set: for modes greater than 2000 cm$^{-1}$ $\Delta$ was set to 200 cm$^{-1}$, and for modes less than 2000 cm$^{-1}$ $\Delta$ was set to 25 cm$^{-1}$.

\subsection{\label{sub:wang}Wang--Landau}
For all Wang--Landau calculations the vibrational quantum number probability increase/decrease parameter was set to 5\%. For the density of states calculation the energy bins were set to 25 cm$^{-1}$. The energy range over which the walks took place was 0--9 eV (0--72,590 cm$^{-1}$). Twenty Wang--Landau iterations of decreasing $f$ were run for each molecule, with each iteration consisting of 1 $\times$ 10$^{7}$ steps.  

The walks to calculate I($\nu$,E) for the absorption spectra were carried out until all of the total internal energy bins were visited an average of 6,000 times. The walks to calculate I$^\prime$($\nu$,E) for the emission spectra were carried out until the energy bins were visited an average of 9,000 times. For the energy--dependent spectra, the accumulated spectrum bins were set to 0.1 cm$^{-1}$. For the high--resolution spectra used in figures \ref{fig:pirali} and \ref{fig:zoom}, the spectrum bins were set to 0.005 cm$^{-1}$.

\subsection{\label{sub:cascade}Cascade spectra}
The spectra of twenty PAHs were simulated using the cascade method described in the previous sections; seven neutral PAHs and their cationic counterparts: naphthalene, anthracene, tetracene, phenanthrene, chrysene, pyrene, and benz[a]anthracene; and six aliphatic containing PAHs: 9--methylanthracene, 9,10--dimethylanthracene, 9,10--dihydroanthracene, 9,10--dihydrophenanthrene, 1,2,3,4--tetrahydronaphthalene, and 1,2,3,6,7,8--hexahydropyrene. For the methylated PAHs, large amplitude motions, such as hindered methyl rotations, were treated at the harmonic level, as we have done previously\cite{mackie2017}. These molecules were chosen based on previous confirmation of their anharmonic IR spectra compared with matrix--isolation data, NIST database gas--phase data, as well as high--resolution, mass--selected, low--temperature, gas--phase spectra\cite{mackie2015b,mackie2016anharmonic,mackie2017}. Simulations were run for each molecule with varying excitation photon energies: 3, 5, 7, and 9 eV. The cascade spectrum of each PAH was simulated using 5 $\times$ 10$^{6}$ UV photons, resulting in approximately 1 $\times$ 10$^{8}$ emitted IR photons per simulation (see supplemental material for individual cascade spectra).

\section{\label{sec:results}Results}
\begin{figure*}
\resizebox{0.90\hsize}{!}{\includegraphics{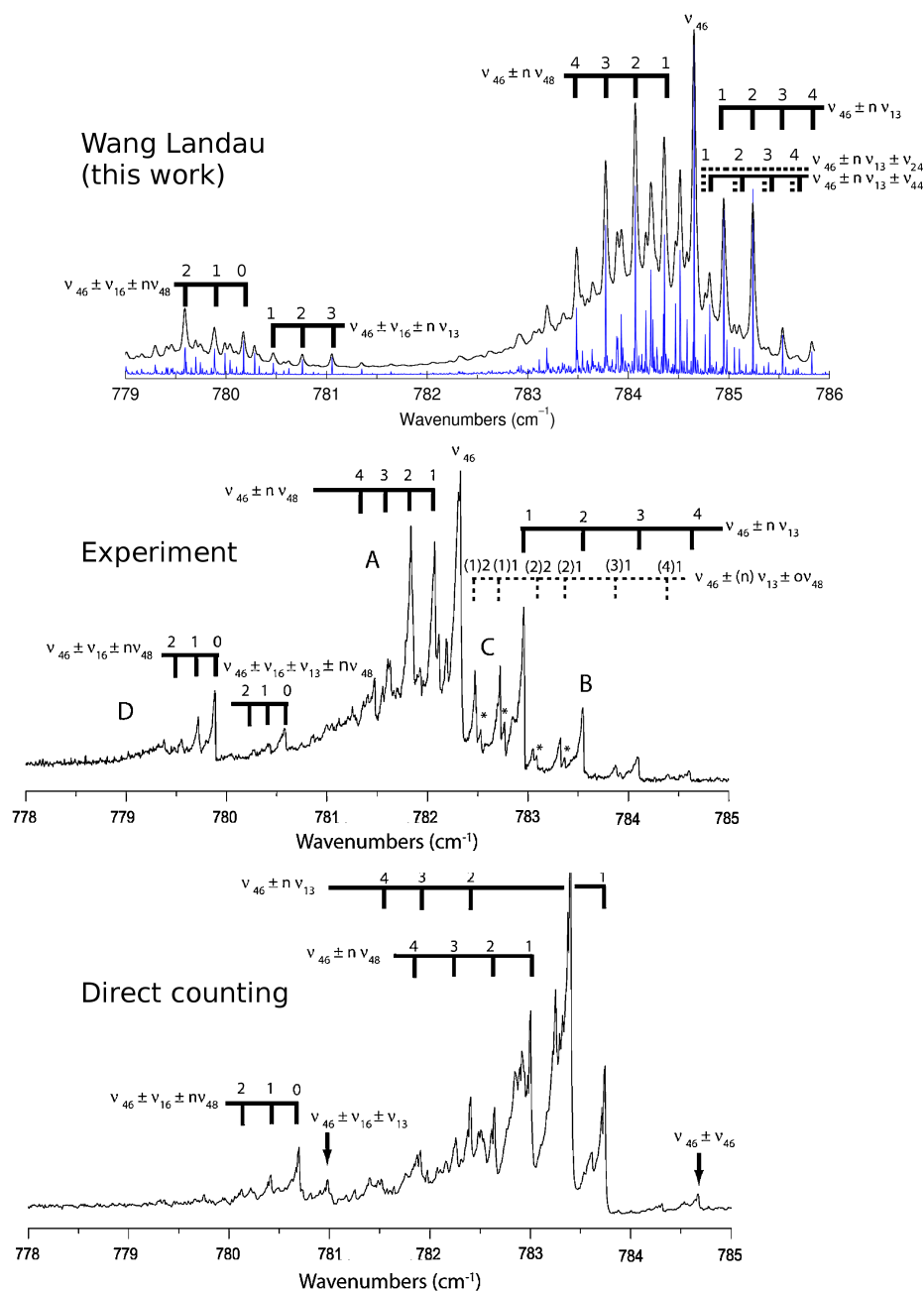}}
\caption{Comparison of high--resolution experimental data of thermally excited naphthalene (300 K)\cite{B814037E} (middle panel) to the ``direct'' theoretical approach\cite{B814037E} (bottom panel) to the Wang--Landau method of this work (top panel, blue, with convolved (HWHM 0.02 cm${^-1}$) spectrum in black). Adapted from Ref. \citenum{B814037E} with permission from the PCCP Owner Societies.}
\label{fig:pirali}
\end{figure*}

\begin{figure*}
\resizebox{0.9\hsize}{!}{\includegraphics{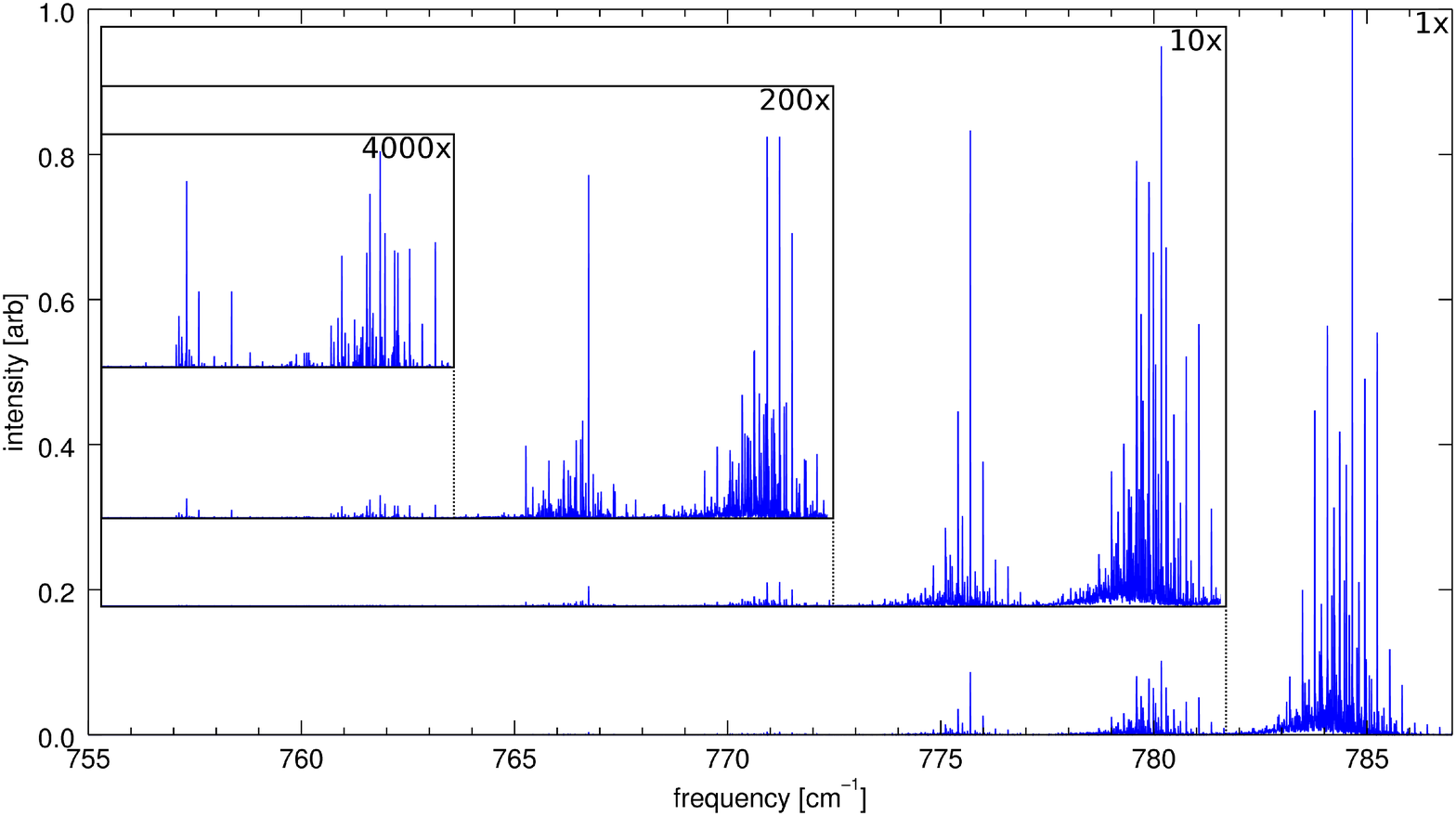}}
\caption{A sampling of the series of band features of the $\nu_{46}$ band of naphthalene at 300 K simulated using the Wang--Landau method for producing a temperature--dependent spectrum. Total magnification is given for each portion of the spectrum.}
\label{fig:zoom}
\end{figure*}

In order to test the reliability of the Wang--Landau method which incorporates polyads, figure \ref{fig:pirali} shows a comparison for the $\nu_{46}$ vibrational band of naphthalene between the Wang--Landau approach, the high--resolution experimental data of reference \citenum{B814037E}, and a direct counting method produce by Pirali et al\cite{B814037E}. The experimental spectrum is dominated by the strong $\nu_{46}$ fundamental, which is accompanied by a large number of well--defined satellite bands characterized by small anharmonic shifts due to one or more excitations in one or more other modes. While the absolute position of the $\nu_{46}$ band is off by approximately 2 cm$^{-1}$ ($\sim$ 0.2\%), the general pattern of main band plus satellite bands is well reproduced both in frequency shifts, number of satellite bands and relative intensities. Pirali et al modeled this spectrum using a direct counting technique which does not separate resonances by symmetry and introduces vibrational population cut--offs to speed up calculations (as described in reference \citenum{B814037E}). This method also results in a dominant $\nu_{46}$ band with a number of satellite bands due to excitation in other modes. However, the pattern is not well reproduced, for example the $\nu_{46} \pm n\nu_{13}$ series is reversed in direction and incorrectly spaced. Agreement in position is better, however this is a result of post--calculation tweaks to the QFF\cite{private}.

The features in the unconvolved spectrum of the Wang--Landau approach of Figure \ref{fig:pirali} (blue) are individual peaks (i.e., not noise). Each feature can be individually assigned with careful (automated) bookkeeping during the energy--dependent spectrum generation step. Figure \ref{fig:zoom} shows an example of a series of magnified panels in order to show the level of detail that can be achieved with the random walk. Since no prior cut--offs of populations are required, the detail becomes limited by the number of Wang--Landau steps taken, and the spectral bin size set initially. A rich progression of bands can be observed.

\begin{figure}
\resizebox{\hsize}{!}{\includegraphics{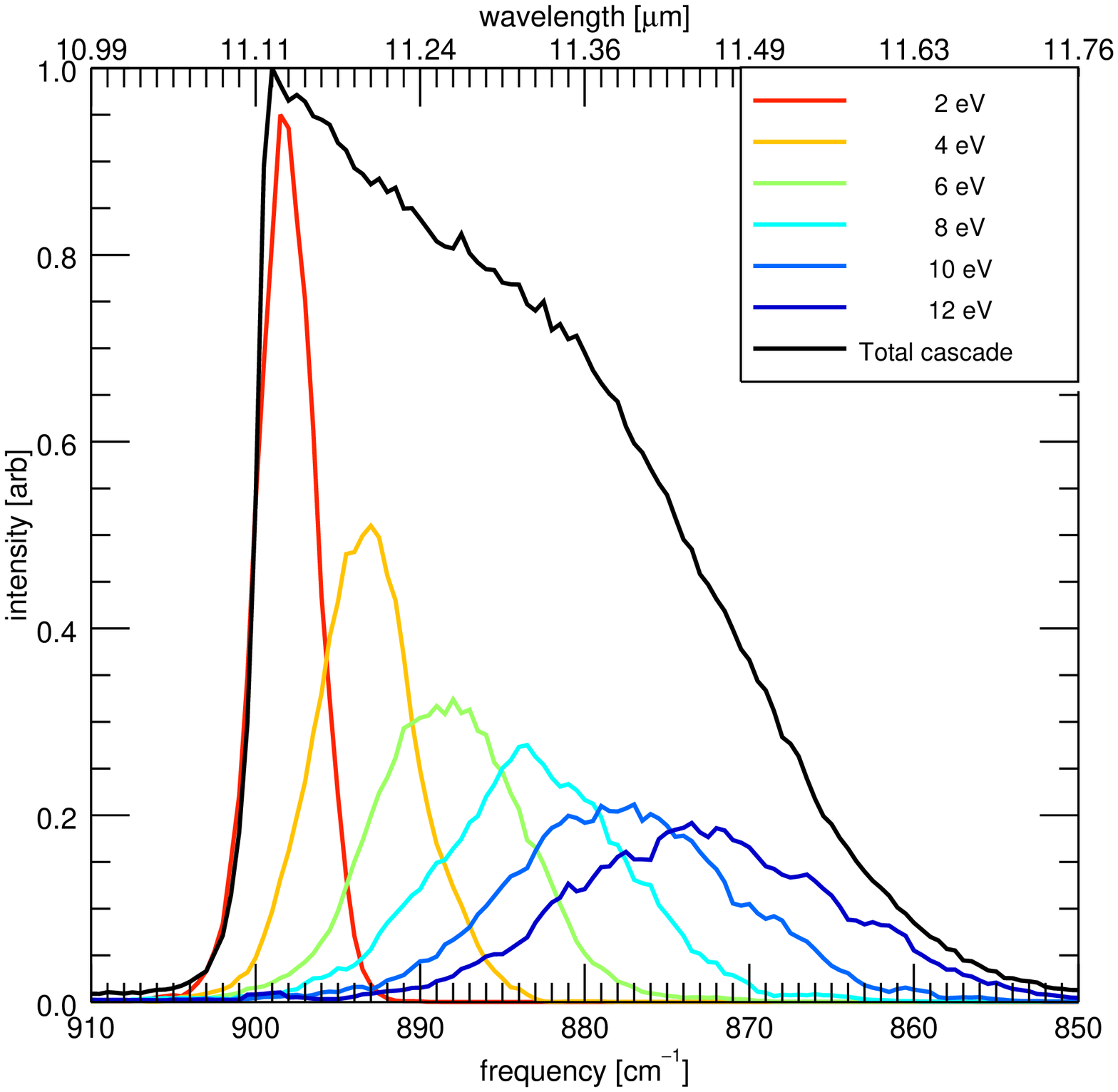}}
\caption{The distribution of IR photon emission probabilities at various internal energies along the cascade of a C--H out--of--plane bending mode of tetracene (colors), with the resulting IR cascade spectrum (black).}
\label{fig:cascade_total}
\end{figure}

\begin{figure}
\resizebox{0.7\hsize}{!}{\includegraphics{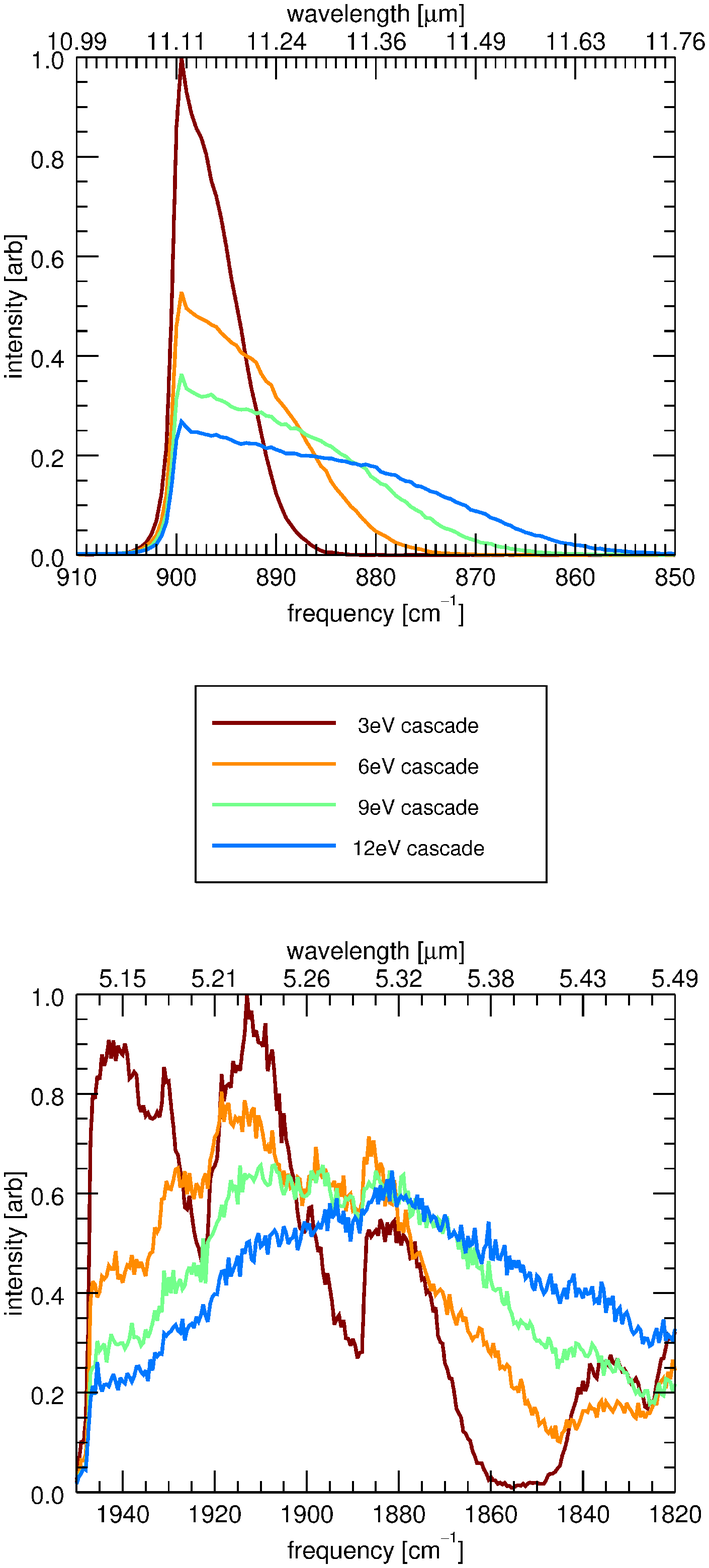}}  
\caption{Example of the pseudoshifts seen in the theoretical IR cascade spectra at various starting cascade energies. Top: an isolated tetracene band, bottom: closely spaced phenanthrene bands.}
\label{fig:shifts}
\end{figure}

\begin{figure}
\resizebox{\hsize}{!}{\includegraphics{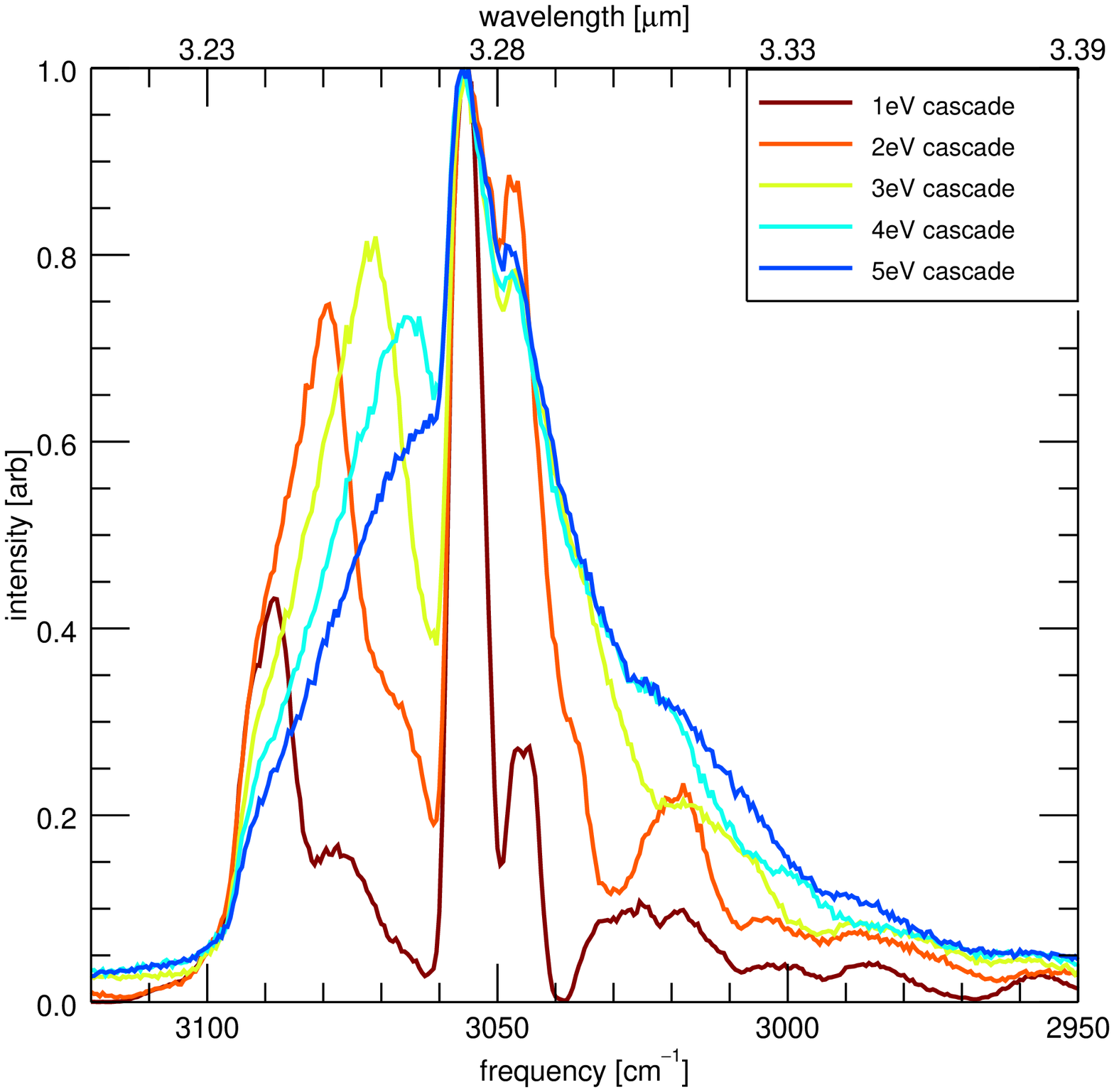}}
\caption{The IR cascade spectrum of the CH--stretching region of tetracene.}
\label{fig:tetracene}
\end{figure}

Large shifts in peak positions are expected for high temperature/highly excited molecules in thermal equilibrium (equation \ref{deltaE}). However, the nature of the IR cascade process leads to total cascade features that are unshifted compared to the ``0 Kelvin'' theoretical anharmonic spectra. This can be reconciled by considering the following: While at any point during the cascade the IR spectrum peak probability of emission is indeed shifted, this emission probability is also broadened. This results in a lower probability of emitting an IR photon at its ``peak'' position when compared to lower internal--energy emission, as shown in Figure \ref{fig:cascade_total}. This results in the piling up of IR photons at relatively the same position during lower internal--energy emissions, while the high internal--energy IR photons are spread over a larger area. An increase in initial UV photon energy therefore results \emph{only} in the growth of pronounced low--energy wings, with no change in the position of the \emph{peak} IR cascade feature. The shifts seen in the resulting spectra are due to secondary effects as described below. As all excited species -- irrespective of their initial starting energy -- go through a low internal energy emission process when the last energy is emitted, a steep blue rise and a red wing is a common characteristic signifying the importance of anharmonicity\cite{1987ApJ...315L..61B, pech2002profiles, joblin2011modeling, cook1998simulated}. No growth is seen in these high--energy wings, leaving an unchanging steep wall on the high--energy side of the majority of features\cite{Candian11042015}.

Shifts of the cascade features \emph{appear} to occur. However, these ``pseudoshifts'' are not due to the typical thermal equilibrium temperature related shifts per se, but are instead due to the cascade process as a whole. The dominant type of pseudoshift occurs when separate IR features are closely spaced in energy. As the initial exciting UV photon energy is increased, more IR photons are released in an ever growing low--energy IR wing. This lowers the relative peak intensity of all of the cascade features, while at the same time if two or more features are close then the low--energy IR wing of one feature slips under the peak of its neighbor. This in turn inhibits the degradation of the intensity of the neighboring feature. This can cause the lower energy positioned feature to become more intense than the previously stronger higher energy positioned feature (as shown in the bottom panel of Figure \ref{fig:shifts}). Even very weak neighboring features can cause the appearance of a shift. The amount of shifting produced by this phenomenon depends on the initial band spacing, and has been observed to occur for bands separated by up to 20 cm$^{-1}$ for cascades starting at high energies. Once the lower intensity neighbor overtakes the initially intense band the gradual appearance of the ``shifting'' stops abruptly. 

Although the vast majority of cascade features do not appear to move with excitation energies, in rare cases the growth of a high energy wing and shift in feature position can occur in the CH--stretching region for combination bands that are of higher frequency than the fundamentals. This occurs due to Fermi resonances. The magnitude of interaction between resonating modes has been shown to change at different internal energies in the CH--stretching region of some PAHs as described in section \ref{sub:polytemp}. Due to differences in the anharmonic constants and quantum occupation numbers, fundamental bands and combination bands shift at different rates, leading to resonances that wax and wane as the states move closer and further apart in energy. This results in more intensity being borrowed by the combination bands at particular internal energies. Although minor in most cases, tetracene shows this effect quite strongly (figure \ref{fig:tetracene}). The feature at 3100 cm$^{-1}$ can be seen to grow and then shrink relative the the strongest band as the cascade starting energy is increased, while the other main features show the typical stationary behavior. 

\section{\label{sec:conclusion}Conclusions}
As shown in this work, a thorough theoretical treatment is necessary to reproduce and understand the IR cascade spectra of PAHs. While previous works have attempted to reproduce the astronomical observations of PAHs, they have exclusively made use of harmonic IR calculations. This limits their applications and trustworthiness. Additionally, with higher spatial and spectral resolution missions coming online, like the James Webb Space Telescope, the known discrepancies between the harmonically derived spectra and experiments can no longer be overlooked. 

It was previously believed that a PAH's IR absorption spectrum greatly differs from an IR cascade emission spectrum. Previous cascade emission models\cite{bauschlicher2008infrared} of PAHs applied a 15 cm$^{-1}$ shift indiscriminately to  all absorption spectra in order to account for the anharmonic emission processes. In thermal equilibrium, spectra do show pronounced anharmonic shifts in the emission peak, in agreement with experiments. However, as this study shows, the peak position of emission bands, in general, does not vary in cascade spectra but rather develops pronounced low frequency wings. In exceptional cases, a pseudoshift may become apparent when a number of bands are closely clustered or when Fermi resonances are particularly important.

While this work shows the complexity necessary to reproduce the temperature/energy--dependent spectra of PAHs, this work also suggests that this complexity may be unnecessary for the cascade spectra. The stability of the peak positions of the cascade spectra enables the use of 0 Kelvin anharmonic theoretical spectra, or low--temperature gas--phase \emph{absorption} experimental spectra to determine the peak positions (or high energy wall) of IR emission cascade spectra. Previous work has suggested this was not possible\cite{brenner1992infrared}. This reinterpretation of the PAH cascade process can be used to both simplify and improve greatly the current astronomical PAH IR cascade spectral models\cite{2014ApJS..211....8B}. To this end, further research into the ``growth'' rate of the cascade wings as a function of initial cascade energy is currently underway.

\section{\label{sec:supp}Supplemental Material}
See supplemental material for the computed IR cascade emission spectra of the individual molecules (listed in section \ref{sub:cascade}) at four initial cascade energies (3, 5, 7, 9 eV), as well as the computed equilibrium geometries and quartic force fields of each molecule.

\acknowledgments
The authors would like to thank the two anonymous reviewers for their helpful comments which improved greatly the manuscript. The authors would also like to thank Xinchuan Huang his consultations and helpful discussions. The spectroscopic study of interstellar PAHs at Leiden Observatory have been supported through a Spinoza award, and through the Dutch Astrochemistry Network funded by the Netherlands Organization for Scientific Research, NWO. This work is partly supported by a Royal Netherlands Academy of Arts and Sciences (KNAW) professor prize. We acknowledge the European Union (EU) and Horizon 2020 funding awarded under the Marie Sklodowska--Curie action to the EUROPAH consortium, grant number 722346. Calculations were carried out on the Dutch national e--infrastructure (Cartesius) with the support of SURF Cooperative, under NWO EW project SH-362-15. TC sincerely thanks the support from Swedish Research Council (grant No. 2015-06501). AC acknowledges NWO for a VENI grant (639.041.543). TJL gratefully acknowledges support from the NASA 12-APRA12-0107 and NASA 16-PDART16 2-0080 grants. This material is based upon work supported by the National Aeronautics and Space Administration through the NASA Astrobiology Institute under Cooperative Agreement Notice NNH13ZDA017C issued through the Science Mission Directorate.

\bibliography{/home/cameron/documents/papers/citations}{}
\bibliographystyle{jcp_astro}

\end{document}